\documentclass[reprint, aip, jcp, amsmath,amssymb, superscriptaddress]{revtex4-1}

\usepackage{graphicx}
\usepackage{dcolumn}
\usepackage{bm}
\usepackage{color, soul}
\usepackage{enumitem}
\usepackage[normalem]{ulem}
\usepackage{comment}
\usepackage[version=4]{mhchem}
\usepackage{breakurl}

\begin{document}
\title{Stoichiometrically-informed symbolic regression for extracting chemical reaction mechanisms from data}
\author{Manuel Palma Banos}
\affiliation{Department of Chemistry, Johns Hopkins University, Baltimore, Maryland, USA}
\author{Joel D. Kress}
\affiliation{Theoretical Division, Los Alamos National Laboratory, Los Alamos, New Mexico, USA}
\author{Rigoberto Hernandez}
\affiliation{Department of Chemistry, Johns Hopkins University, Baltimore, Maryland, USA}
\author{Galen T. Craven}
\affiliation{Theoretical Division, Los Alamos National Laboratory, Los Alamos, New Mexico, USA}

\begin{abstract}
A data-driven computational method is introduced to extract chemical reaction mechanisms from time series chemical concentration data. 
It is realized through the use of dynamic symbolic regression in which a sparse analytical form for a dynamical system is discovered from the underlying data. 
We specifically develop the stoichiometrically-informed symbolic regression (SISR)
method to address a standing challenge in complex chemical reaction networks:
Given a time-series dataset of concentrations of several components,
what is the mechanism and the associated rate constants?
SISR finds the optimal mechanism, 
kinetic equations and rate constants by combining
 differential optimization with a genetic optimization approach that searches a symbolic space of possible reaction mechanisms.
Use of SISR in several paradigmatic examples spanning linear and nonlinear reaction schemes
results in excellent agreement between true and predicted mechanisms.
The advantages of a stoichiometrically-informed approach
such as SISR to address reaction discovery is illustrated 
through comparison with the use of generic state-of-the-art data-driven approaches.
\end{abstract}

\maketitle

\section{Introduction}

Determining chemical reaction mechanisms is foundational in many research areas such as catalysis, electrochemistry, combustion, and biochemistry 
that feature prominently in modern scientific and technological landscapes \cite{Harper2011comprehensive,Zhu2005modeling,Gossler2019catalytic,Abramovitch2024cancer,Stocker2020machine,Bures2023organic,Savoie2021simultaneously,SavoiePNAS2023, Komatsuzaki2024, Komatsuzaki2025}. 
Chemical reaction mechanisms give fundamental insight into a physicochemical process, providing elucidation and allowing interpretation of the underlying chemical reactions that give rise to a process. Chemical mechanisms can also be used to forecast how the outcome or output of a process will change over time.
However, deriving a set of kinetic mechanistic equations that accurately describes the time evolution of concentrations of chemical species involved in a mechanism is often difficult or simply intractable in practice due to, for example, complex nonlinear interactions between reacting species, a large number of chemical species participating in the process, and/or 
reactions occurring over multiple timescales.
Deriving chemical reaction mechanisms by hand generally requires physical intuition about a system and subject matter expertise \cite{Liu2004DNA,YangKress2023}.
This is because determining accurate functional forms for reaction mechanisms typically involves searching a vast space of possible reactions that are involved in a process while also determining how those reactions are coupled in the overall mechanism.
This problem is compounded because not only
does the reaction mechanism itself need to be determined, but the chemical rate constants describing 
species-to-species transformations, i.e., chemical reactions,  must be parameterized, often over varying thermodynamic conditions such as different temperatures and pressures \cite{Pollak2005reaction,Bazant2013,craven15c,craven16c,matyushov16c}.

Because of these difficulties, automated reaction mechanism generation is an emerging data-driven research approach that can accelerate the extraction of accurate reaction mechanisms from data \cite{Stocker2020machine,Bures2023organic,Savoie2021simultaneously,SavoiePNAS2023, Wilary2023, Carvalho2024, Huang2022}.
Data-driven and machine learning (ML) methods have been broadly and successfully applied in many areas of the physical sciences \cite{carleo2019machine,zhong2021machine,tarca2007machine,wang2019machine,welborn2018transferability,Kulichenko2021review,butler2018machine,Carrasquilla2017,Carleo2017,johnson2024machine,Biamonte2017quantum,Deng2017,Liu2019,craven20b,craven20c}.
Data-driven approaches for reaction discovery have been used to decipher complex and large datasets of chemical concentration data
by extracting chemical reaction pathways, rate constants, and reaction mechanisms \cite{Stocker2020machine,Bures2023organic,Savoie2021simultaneously,Jiang2022identification,SavoiePNAS2023}.
The current data-driven reaction mechanism discovery methods, however, generally suffer from the same shortcomings that are typical of most data-driven and ML approaches including lacking interpretability, a large number of parameters, the black-box nature of the approximating function, and poor performance when extrapolation outside of the training data is performed.
One ML approach that is used to circumvent these limitations is Symbolic Regression (SR)---a method to search for simple analytical functions that best describe a dataset. SR has been applied in multiple contexts 
to extract sparse and interpretable functional forms from data \cite{TegmarkAIFeynmann, Angelis2023artificialSR, Wang2019SRmaterials}.
In the context of SR applied to dynamical systems \cite{Brunton2016SINDY},
dynamical SR approaches can be used to extract a sparse analytical form for a dynamical system  from time series data.
Integrating the discovered system of dynamical equations will generate the time-evolution of the input variables, for example the time-dependence of the concentrations of chemical species \cite{ReactiveSINDY,Bhatt2023SINDYCRN}.

One of the most prominent SR methods that is used to discover
the dynamical equations giving rise to a time series dataset is the Sparse Identification of Nonlinear Dynamical systems (SINDy) approach\cite{Brunton2016SINDY,Kaheman2020SINDYPI,DeSilva2020pysindy}.
SINDy has been applied broadly in the physical sciences and has seen successful applications in diverse areas such as
biological networks \cite{Prokop2024biological,Sandoz2023SINDy, Prokop2024chaos},  aerodynamics \cite{Li2019timevaryingSINDY}, and plasma physics \cite{Dam2017SINDyplasma},
among others.
One of the primary advantages of SINDy is that given a collection of time series data, it can quickly (relative to, for example, genetic SR approaches \cite{Sakamoto2001, Quade2016SR}) search a large space of possible analytical functional forms and corresponding parameter values to generate a simple dynamical system that when integrated matches the time evolution of the input data. 
Other approaches to determine symbolic expressions for physical mechanisms have been developed. For example, the 
sure-independence screening and sparsifying operator (SISSO) method \cite{SISSO2018,Purcell2023sisso}
can be applied to generate sparse analytical descriptors of a material's properties.

The SINDy framework has been applied to determine chemical reaction mechanisms using an approach termed Reactive SINDy (see Ref.~\citenum{ReactiveSINDY}).
This interesting application of SR to chemical reaction networks
produced sparse reaction mechanisms in good agreement with the input data
and illustrates the potential of dynamical SR in macroscale chemical reaction dynamics. 
The Reactive SINDy approach does have limitiations that reduce its utility and robustness including: (1) The user must propose a collection of reaction ansatz, i.e., the user must guess what reactions are present in a process. This requirement can be cumbersome and time-consuming especially if the number of species being studied is large or little is known about the physical process being examined. (2) There are no constraints on the reaction rates in the derived mechanism which can lead to unphysical results such as negative concentrations when the system is integrated. (3) Fast-slow dynamics \cite{Bramburger2020timescale} are not well-described. For example, if there are rate constants that differ by multiple orders of magnitude the SINDy approach will generally prune the slow process from the derived dynamical system. This poses a problem because elimination of reactions with small rate constants can eliminate reaction pathways that are vital to the overall mechanism.
While these problems are not present in all chemical reaction networks/mechanisms, they do limit the 
applicability and utility of Reactive SINDy in some cases \cite{ReactiveSINDY}.
Other approaches such as  SINDy - CRN \cite{Bhatt2023SINDYCRN} where CRN stands for Chemical Reaction Network, and the one defined in Ref. ~\citenum{Ducci2025} seek to alleviate some of these problems.

Other methodologies for extracting dynamical systems from data have been developed and applied to good effect \cite{Wilson2019AutomatedCRN, Chen2018KineticSpectraUnknownAbsorbers, Bradley2022NeuralDEStiff, Gusmao2023KINN, Prabhu2025DFDIDiscovery, Wu2022PolyODENet, Muthyala2025SyMANTIC, Lee2025separations}. 
In the context of chemical reaction mechanism discovery, it would be advantageous to developing a method that (a) gives the explicit individual reactions involved in a process,
(b) gives the stoichiometry of those reactions---a fundamental property in the analysis of chemistry and chemical reactions, 
(c) does not rely on neural network formulations of the chemical reaction network, as they can reduce interpretability and accurate extrapolation (i.e., accurate time-series forecasting), 
(d) can detect hidden variables such as unknown chemical intermediates in a reaction mechanism, 
(e) does not require a postulated set of potential reactions be included as reaction ansatz,
(f) accurately returns the rate constants for each reaction,
and (g) is robust to noise in data.

In this work, we develop and apply a stoichiometrically informed symbolic regression (SISR---pronounced ``scissor'')  tool to determine chemical reaction mechanisms and chemical kinetic equations from time-series concentration data.
Our specific technical advance is to apply a physics-informed mathematical formalism that accounts for intrinsic stoichiometry in a chemical reaction to automate the discovery of accurate chemical reaction mechanisms from data. The developed method returns sparse and interpretable analytical forms for a reaction mechanism discovered from data. A genetic optimization approach is employed to search the symbolic space of possible reaction mechanisms to find the one that best matches a time-series dataset of chemical concentrations. 
That genetic approach is coupled with the stoichiometrically-informed method to fit the rate constants in a reaction mechanism through differential optimization. Applying the method results in excellent agreement between true and predicted mechanisms over data from multiple linear and nonlinear reaction schemes. The agreement is shown to persist over sparse and noisy datasets, such as those that would typically be obtained from experiments.

The remainder of this article is organized as follows:
Section~\ref{sec:methods} contains details of methods and the formalism that are applied including the genetic search procedure over the symbolic reaction space
and the numerical procedures used to fit the rate constants in those reactions. 
In Sec.~\ref{sec:results}, the results of the method for several model reactive schemes and chemical reaction networks are presented.   
Conclusions and future directions are discussed in Sec.~\ref{sec:conc}.

\section{Stoichiometrically-Informed Symbolic Regression (SISR) \label{sec:methods}}

\begin{figure*}
 \centering
 \includegraphics[width=18.2cm]{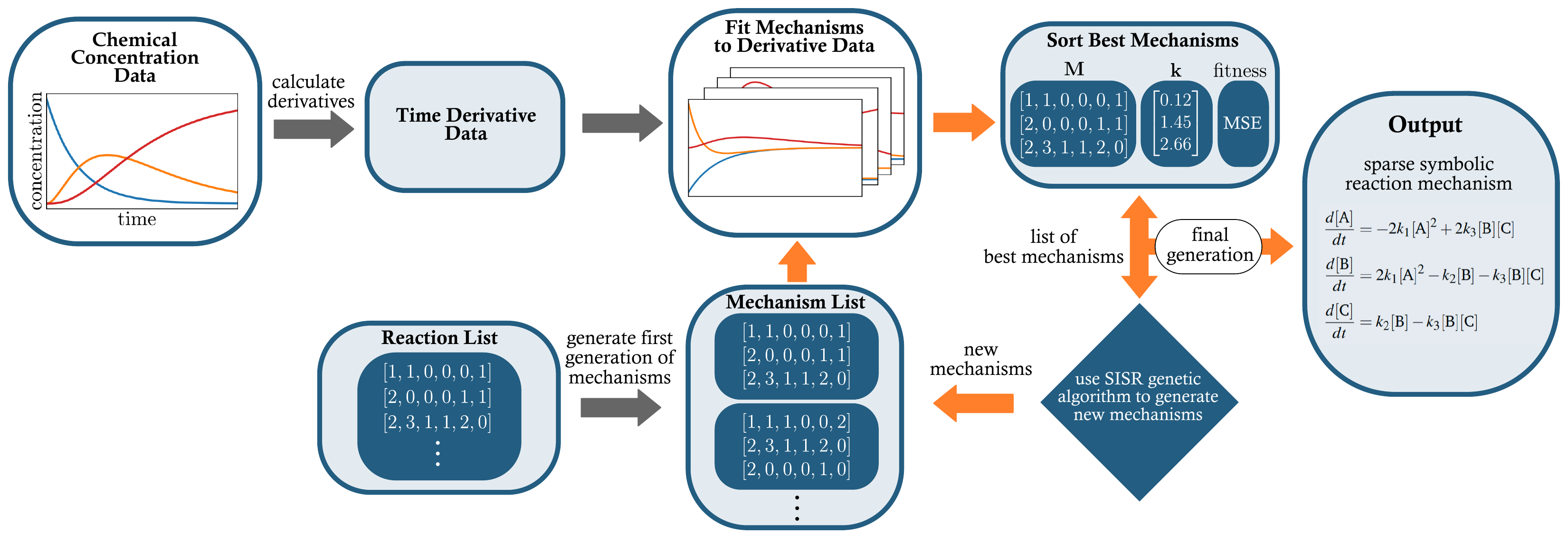}
 \caption{Schematic diagram showing the workflow for the developed SISR method.}
 \label{fig:workflow}
\end{figure*}

\subsection{Data Structure and Mathematical Formalism}

The overall goal of SISR is to take time series concentration data for chemical processes where the underlying reactions are unknown, and to extract the correct reactions and rate constants from that data. 
The SISR method is described using the following mathematical formalism. 
Consider a dataset of chemical concentration data containing $N$ chemical species, where the concentration of each species is measured at times $t_1,t_2, \ldots, t_m$. Expressed in matrix
form where each column is the times series concentration data for a different species, this dataset is
\[
\text{S} = \begin{bmatrix}
  [\text{S}_1](t_1) & [\text{S}_2](t_1) & \cdots & [\text{S}_N](t_1) \\
  [\text{S}_1](t_2) & [\text{S}_2](t_2) & \cdots & [\text{S}_N](t_2) \\
  \vdots        & \vdots        & \ddots & \vdots        \\
  [\text{S}_1](t_m) & [\text{S}_2](t_m) & \cdots & [\text{S}_N](t_m)
\end{bmatrix}.
\]
where the $[\text{S}_k]$ notation represents the concentration of chemical species $\text{S}_k$.
The goal is to derive a {\it symbolic} reaction mechanism $\bold{M}$ and the corresponding set of rate constants $\bold{k}$
that best fits the dataset $\text{S}$.
The total reaction mechanism comprises a set of chemical reactions involved in the process, the rate constants for those reactions, and the corresponding  set of kinetic equations for the set of reactions.
Our approach is to use a genetic algorithm to evolve stoichiometry-constrained symbolic expressions 
for a collection of reaction mechanisms, fit the rate constants in those mechanisms to numerical derivatives of the concentration data, and repeat this process for multiple iterations (generations) using the best mechanisms from the previous generation to generate the next generation. After a set number of generations are evolved,  a
final determination of the best overall mechanism is made. A schematic diagram of the developed workflow is shown in Fig.~\ref{fig:workflow}.

The mechanisms are fit and constructed in the derivative space of the concentration data.
Here, the numerical derivatives of the concentration
data, obtained using finite difference methods, are represented by
\[
\dot{\text{S}} = \begin{bmatrix}
  \frac{d[\text{S}_1]}{dt}\Big|_{t = t_1} & \frac{d[\text{S}_2]}{dt}\Big|_{t = t_1} & \cdots & \frac{d[\text{S}_N]}{dt}\Big|_{t = t_1} \\
  \frac{d[\text{S}_1]}{dt}\Big|_{t = t_2} & \frac{d[\text{S}_2]}{dt}\Big|_{t = t_2}& \cdots & \frac{d[\text{S}_N]}{dt}\Big|_{t = t_2} \\
  \vdots        & \vdots        & \ddots & \vdots        \\
  \frac{d[\text{S}_1]}{dt}\Big|_{t = t_m} & \frac{d[\text{S}_2]}{dt}\Big|_{t = t_m} & \cdots & \frac{d[\text{S}_N]}{dt}\Big|_{t = t_m} 
\end{bmatrix}.
\]
Fitting to derivatives is a common approach used to discover symbolic dynamical systems from data because it allows
the derivative functions, for example, $\frac{dx}{dt}$, $\frac{dy}{dt}$, $\frac{dz}{dt}$, to be constructed directly as opposed to constructing the primary 
functions, $x(t)$, $y(t)$, $z(t)$, and then deriving the dynamical system from those functions \cite{Brunton2016SINDY, de2020pysindy}. We apply this same approach.

A symbolic reaction mechanism $\bold{M}$ is a collection of chemical reactions 
\begin{equation}
\bold{M} = \left[\text{rxn}_i,\text{rxn}_j, \text{rxn}_k, \ldots\right],
\end{equation}
where each reaction ($\text{rxn}$) is associated with a rate constant $k_\text{rxn}$.
The total mechanism is a combination of the symbolic mechanism $\bold{M}$ and an array of rate constants $\bold{k}$ that contains a numerical value for the rate constant of each  reaction in that mechanism.
For example, some reactions that are possible in a process involving three chemical species
$\text{S}_1 = \text{A}$, $\text{S}_2 = \text{B}$,and $\text{S}_3 = \text{C}$
are
\begin{equation}
\label{eq:example1}
\begin{aligned}
\ce{2 A &->  B }, \\[0ex]
 \ce{A + B &->  C },\\[0ex]
\ce{2 A + 3 B + C &-> A + 2 B }.
\end{aligned}
\end{equation}
The size of a mechanism is given by its cardinality $|\bold{M}|$ which is the number of reactions in the mechanism.

To illustrate the SISR method, consider the abstract chemical reaction example
\begin{equation}
\label{eq:abstractrxn}
\ce{$\sum_{i=1}^N$ $s^{(\text{r})}_{i}$ S_$i$  ->[$k$] $\sum_{i=1}^N$ $s^{(\text{p})}_i$ S_$i$ } \\[-1ex]
\end{equation}
where the terms on the LHS are reactants (denoted by the superscript ``$\text{r}$'') and the terms on the RHS are products (denoted by the superscript ``$\text{p}$'').
The sums are taken over all the chemical species involved in the process.
The stoichiometric coefficients of species $\text{S}_i$ in the reactant and product states are $s^{(\text{r})}_i$  and $s^{(\text{p})}_i$, respectively.
To represent each reaction in vector form (a convenient mathematical notation for our purposes), 
we break each reaction into a reactant vector and a product vectors, each containing the stoichiometric coefficients
in the respective state.
The reactant vector for Eq.~(\ref{eq:abstractrxn}) is 
\begin{equation}
\text{rxn}^{(\text{r})} = \left[s^{(\text{r})}_1,s^{(\text{r})}_2,\ldots,s^{(\text{r})}_N\right],
\end{equation}
the product vector is
\begin{equation}
\text{rxn}^{(\text{p})} = \left[s^{(\text{p})}_1,s^{(\text{p})}_2,\ldots,s^{(\text{p})}_N\right],
\end{equation}
and the total reaction vector is obtained through a concatenation of the reactant and product vectors:
\begin{equation}
\begin{aligned}
\label{eq:rxnvec}
\text{rxn} &=  \left[s^{(\text{r})}_1,s^{(\text{r})}_2,\ldots,s^{(\text{r})}_N\right]  \bigoplus \left[s^{(\text{p})}_1,s^{(\text{p})}_2,\ldots,s^{(\text{p})}_N\right] \\ 
&= \left[s^{(\text{r})}_1,s^{(\text{r})}_2,\ldots,s^{(\text{r})}_N, s^{(\text{p})}_1,s^{(\text{p})}_2,\ldots,s^{(\text{p})}_N \right] \in \mathbb{N}_0^{2N},
\end{aligned}
\end{equation}
where $\oplus$ represents the concatenation operation.
For example, consider again a reaction involving three species: $\text{S}_1 = \text{A}$, $\text{S}_2 = \text{B}$,and $\text{S}_3 = \text{C}$
defined by
\begin{equation}
\ce{2 A + 3 B + C ->[$k$]  A + 2 B }
\end{equation}
The reactant vector for this reaction is
\begin{equation}
\text{rxn}^{(\text{r})} = [2,3,1],
\end{equation}
the product vector is
\begin{equation}
\text{rxn}^{(\text{p})}  = [1,2,0],
\end{equation}
and the total reaction vector is 
\begin{equation}
\text{rxn} =  [2,3,1]  \bigoplus [1,2,0]  = [2,3,1,1,2,0].
\end{equation}
Using this formalism, chemical reactions can be expressed in convenient mathematical form that maps to a symbolic representation for analysis and optimization.

Each total mechanism has a corresponding system of symbolic kinetic equations defined by:  
\begin{equation}
\begin{aligned}
\label{eq:kinetic}
\frac{d[\text{S}_1]}{dt} &=   \sum_{\text{rxn} \in \textbf{M}} \! k_\text{rxn} \left(s^{(\text{p})}_1 - s^{(\text{r})}_1\right) \prod^N_{j=1}[\text{S}_{j}]^{s^{(\text{r})}_j},   \\[-1ex]
\frac{d[\text{S}_2]}{dt} &=  \sum_{\text{rxn} \in \textbf{M}} \! k_\text{rxn} \left(s^{(\text{p})}_2 - s^{(\text{r})}_2\right) \prod^N_{j=1}[\text{S}_{j}]^{s^{(\text{r})}_j}, \\[-1ex]
 &\cdots& \\[-1ex]
\frac{d[\text{S}_N]}{dt} &=  \sum_{\text{rxn} \in \textbf{M}} \! k_\text{rxn} \left(s^{(\text{p})}_N - s^{(\text{r})}_N\right) \prod^N_{j=1}[\text{S}_{j}]^{s^{(\text{r})}_j}. 
\end{aligned}
\end{equation}
The kinetic equations are integrated to determine the time evolution of the concentration of each species.

\subsection{Genetic Search for the Optimal Reaction Mechanism}

We now describe the mathematics of how symbolic mechanisms are evolved in SISR. 

\subsubsection{Reaction List}
The first step is to create a reaction list---a list of all possible reactions that could be included in a mechanism.
The reaction list is created based on predefined constraints for the \textit{maximum reaction order}, $O$, and  \textit{maximum stoichiometric ratio}, $R$, that can be included in each reaction on the list.
Each reaction in a mechanism is represented by a reaction vector like in Eq.~(\ref{eq:rxnvec}) with $2N$ elements, where the first $N$ elements are the stoichiometric coefficients of the reactants, the sum of which must satisfy the constraint
\begin{equation}
\sum_{i=1}^{N} s^{(\text{r})}_i  \leq O.
\end{equation}
This is a mathematical statement that a reaction on the reaction list must not exceed the predefined reaction order. Typical values for $O$ will be $O=1$ if
only first order (and optionally zeroth order) reactions are to be included or $O=2$ if first and second order reactions (and optionally zeroth order) are to be included, although
higher reaction orders can also be used. 
The last $N$ elements in the reaction vector are the stoichiometric coefficients of the products, the sum of which in the SISR method is constrained by:
\begin{equation}
\sum_{i=1}^{N} s^{(\text{p})}_i  \leq O*R.
\end{equation}
The number of possible reactant vectors is given by $N+O$ Choose $O$, i.e.,  $\binom{N+O}{O}$, and the number of possible product vectors is $(O*R)^{N}$.
The final list of reactions is formed by combining every possible reactant vector with all possible  product vectors and eliminating redundant or stoichiometrically prohibited reactions.

We utilize an \textit{islanding} procedure in the SISR genetic algorithm, where multiple islands are created, each containing mechanisms with a fixed number of reactions, $|\bold{M}|$. Islanding is a technique used in genetic algorithms where the population is divided into distinct subpopulations (here based on the number of reactions in a mechanism), each evolving separately on different "islands". The mechanisms in each island maintain a constant number of reactions throughout the evolutionary process. The genetic algorithm is then applied separately to each island, allowing independent evolution of mechanisms with their respective constraints on the number of reactions. No information is transferred between islands during the search procedure. The islanding approach promotes diversity among potential solutions and mitigates the need to address mechanism complexity (or other factors) for optimal mechanism selection during the search on each island.
This ultimately leads to more efficient optimization and also makes the search easier to parallelize.

\subsubsection{First generation}

In the next step, the first generation of possible reaction mechanisms 
\begin{equation}
\textbf{G}_1 =   \left[\bold{M}_1, \bold{M}_2, \bold{M}_3, \ldots \right],
\end{equation}
is constructed for each island using the reaction list.
In the first generation, $n_\text{mech}$ mechanisms are constructed. Each mechanism is constructed by randomly and sequentially selecting reactions from the reaction list until (a) all the chemical species that are involved in a process are included in that mechanism (either as reactant, product, or both) and (b) the number of reactions in the mechanism is equal to the cardinality of the specific island size.
The first constraint is imposed so that if the concentration data set contains data for $N$ chemical species, then every mechanism should involve $N$ chemical species, i.e., if $N$ chemical species are represented in the data, then all $N$ species must be involved in the mechanism. 
Each reaction in the reaction list is equally weighted in this initial selection procedure.
The outcome is collection of $n_\text{mech}$ possible mechanisms.

After constructing symbolic expressions for each reaction mechanisms in the generation, the next step is to fit the rate constants for each mechanism to the data. Our aim is to minimize the discrepancy between the true concentration derivatives $\frac{d[\text{S}_i]}{dt}$ and the predicted derivatives $\frac{d[\hat{\text{S}}_i]}{dt}$ for each species $i$ where the hat notation signifies that the derivatives arise from fits to the data.
The time derivatives are computed using a second-order accurate central difference for interior points and first-order accurate forward/backward differences at the boundaries. This approach is able to handle both uniformly-spaced and nonuniformly-spaced time-series datasets.    
The rate constant fitting is achieved using the mean squared error (MSE) defined through the loss function in the derivative space
\begin{equation}
\label{eq:MSE}
\mathcal{L}_\text{der} \equiv \frac{1}{Nm}\sum^m_{j = 1} \sum^N_{i =1} \left(\dfrac{ \frac{d[\text{S}_i]}{dt}\Big|_{t = t_j}}{ \text{max}\left(\Big|\frac{d[\text{S}_i]}{dt}\Big|\right)} - \dfrac{\frac{d[\hat{\text{S}}_i]}{dt}\Big|_{t = t_j}}{\text{max}\left(\Big|\frac{d[\hat{\text{S}}_i]}{dt}\Big|\right)}\right)^2,
\end{equation}
as an overall loss metric for each mechanism where the sum over $j$ accounts for all the time points in the dataset and the sum over $i$ accounts for all the chemical species.
Each derivative value in the MSE is scaled so that species with large derivative value do not dominate the error calculation.
The fitting process is performed by finding the set of rate constants that minimize the MSE between the observed and predicted derivatives: 
\begin{equation}
\mathbf{k}_{\text{fit}} = \underset{\mathbf{k}}{\mathrm{argmin}} (\mathcal{L}_\text{der}(\mathbf{k})).
\end{equation}
The time derivatives for each species in a mechanism are fit to expressions that encode stoichiometric information in the form: 
\begin{equation}
\frac{d[\hat{\text{S}}_i]}{dt}\Big|_{t = t_j}  = \sum_{\text{rxn} \in \textbf{M}} \! k_\text{rxn} \left(s^{(\text{p})}_i - s^{(\text{r})}_i\right) \prod^N_{j=1}[\text{S}_{j}](t_j)^{s^{(\text{r})}_j},
\end{equation}
for each species.
The minimization is performed using nonlinear least squares regression implemented through the trust region reflective method.
All of the rate constants are fit at the same time, as opposed to a sequential fitting procedure.
After numerical values for the rate constants in every mechanism have been assigned using the fitting procedure, we have the first generation of fit mechanisms 
\begin{equation}
\textbf{G}^{(\text{fit})}_1 =   \left[\bold{M}^{(\text{fit})}_1, \bold{M}^{(\text{fit})}_2, \bold{M}^{(\text{fit})}_3, \ldots \right],
\end{equation}
and a corresponding set of values for the loss function
\begin{equation}
\label{eq:errorarray}
\mathcal{L}_1 =   \left[\mathcal{L}^{(1)}_\text{der}, \mathcal{L}^{(2)}_\text{der}, \mathcal{L}^{(3)}_\text{der}, \ldots \right].
\end{equation}
The initial generation of mechanisms is then sorted and ranked based on fitness defined by the MSE in the derivative space, i.e., $\mathcal{L}_\text{der}$.

\subsubsection{Next generations}

The  $n_\text{best} = \text{int}(\mathcal{E}n_\text{mech})$ fittest mechanisms from the previous generation $\textbf{G}_{i}$  are kept for the next generation $\textbf{G}_{i+1}$ using an elitism fraction $\mathcal{E}$.
Therefore, in each subsequent generation after the first, $n_\text{new} = n_\text{mech} - n_\text{best}$ new mechanisms must be created using
information from the best mechanisms from the previous generation.
This process is performed using crossover methods and then mutation methods.

Crossover involves taking information (reactions, reactant vectors, and/or product vectors) from the best performing mechanisms and using 
that information to generate new mechanisms. 
The probability of a mechanism from the previous generation being involved in a crossover event comes from the ranked-based selection:
\begin{equation}
    p_{i} = \frac{n_\text{mech}-i + \epsilon}{\displaystyle \sum_{j=1}^{n_\text{mech}} (n_\text{mech}-j + \epsilon)},
\end{equation}
where $\epsilon$ is a small fractional numerical value (throughout taken to be 0.2) used so that a non-zero probability is 
assigned to all the mechanisms in a generation.
We use this ranking procedure instead of weighting directly according MSE value to avoid high-fitness mechanisms dominating in early generations. 
The ranked-based weighting adds diversity to the pool of solutions.

The crossover mechanism is generated by choosing two mechanisms (the parents) randomly according to the probability $p_{i}$.
The reactions contained in the two selected parent mechanisms are then combined into a \textit{gene pool}
which is a collection of all the reactions involved in the chosen mechanisms. 
The gene pool is edited so reactions only appear once.
Two offspring mechanisms are then created by randomly selecting reactions from this \textit{gene pool}.
The offspring are created by randomly and sequentially selecting reactions from the gene pool until all the chemical species that are involved in a process are included in children mechanisms and the number of reactions in the mechanism is equal to the cardinality of the specific island size being evolved.
For example, consider the two parent mechanisms from an $|\bold{M}| = 4$ island:
\begin{equation}
\bold{M}_1 = \left[\text{rxn}_1,\text{rxn}_2, \text{rxn}_3, \text{rxn}_4 \right],
\end{equation}
\begin{equation}
\bold{M}_2 = \left[\text{rxn}_5,\text{rxn}_6, \text{rxn}_7, \text{rxn}_8 \right].
\end{equation}
The gene pool for these parents is
\begin{equation}
\text{GP} = \left[\text{rxn}_1,\text{rxn}_2, \text{rxn}_3, \text{rxn}_4, \text{rxn}_5,\text{rxn}_6, \text{rxn}_7, \text{rxn}_8  \right].
\end{equation}
Using this gene pool to generate two children, results in mechanisms such as:
\begin{equation}
\bold{M}_{\text{child}_1} = \left[\text{rxn}_1,\text{rxn}_2, \text{rxn}_5, \text{rxn}_6 \right].
\end{equation}
and
\begin{equation}
\bold{M}_{\text{child}_2} = \left[\text{rxn}_1,\text{rxn}_5, \text{rxn}_7, \text{rxn}_8 \right].
\end{equation}
Crossover can be chosen to occur over specific reactions as described above, or in the reactant and/or product vectors separately.
The latter will be advantageous when examining chemical processes with large number of chemical species.

Once the new generation of mechanisms is created using crossover, a random number of the newly-generated mechanisms are selected to be mutated. 
Mutation is not performed on the $n_\text{best}$ elite mechanisms, only on the $n_\text{new}$ new mechanisms generated using crossover.
This is to retain the best solutions, otherwise the mutation could take an optimal solution and change it, removing important reaction information from the overall gene pool.
Throughout this work we use a mutation rate $\mathcal{M}$ of 0.1, meaning $10\%$ of the new mechanisms are mutated.
In the mutation procedure, first, a random mechanism is selected from $n_\text{new}$ mechanisms. Next, a random reaction is selected from the selected mechanism and is substituted for another reaction from the original reaction list, with all reactions on this list being equally probable. 
Finally, the new mechanism with the substituted reaction is
checked to see if it satisfies the constraint
that all chemical species in the dataset are included. If so,
the new mechanism is substituted with original mechanism before mutation.
If not, then the mutation procedure starts over by selecting a new mechanism to mutate, and the original mechanism stays on the list of mechanisms.

The rate constants in the new generation are then fit to constrict the new generation of fit mechanisms $\textbf{G}^{(\text{fit})}_{i+1}$ and a corresponding set of values for the loss function $\mathcal{L}_{i+1}$.
Then, the sort $\to$ crossover $\to$ mutate algorithm starts again until a set number of generations are evolved.

\subsubsection{Final generation and mechanism selection}

Because islanding is used in the SISR genetic algorithm, 
after the final generation of mechanisms is generated there is not a single best solution but
instead a collection of best solutions, one for each island. 
Therefore, final selection of the overall best mechanism must be made.
There are several ways to approach this problem.
Here, we employ an approach based on multiobjective optimization.
Specifically, we 
seek the mechanism that minimizes the discrepancy (the MSE) between the scaled ground truth concentration data and the data generated by 
the extracted mechanism:
\begin{equation}
\label{eq:Lc}
\mathcal{L}_\text{c}\equiv\frac{1}{N m}\sum^m_{j = 1} \sum^N_{i =1} \left( \frac{[\hat{\text{S}}_i](t_j)}{\text{max}\left([\hat{\text{S}}_i]\right)} 
-  \frac{[\text{S}_i](t_j)}{\text{max}\left([\text{S}_i]\right)} 
\right)^2,
\end{equation}
and that also minimizes the complexity of the derived mechanism \cite{note1}. 
Complexity metrics in symbolic regression are used to quantify the simplicity or sparsity of the generated models. These metrics penalize overly complex expressions that do not significantly improve predictive performance \cite{Kommenda_2015, haut2025data, smits2005pareto, Ducci2025}.
Notice in Eq.~(\ref{eq:Lc}) that while we discover the total mechanisms in the derivative space, the final determination of the best mechanism is made based on the concentration error, which is then coupled with a complexity metric.

\begin{figure}
 \centering
 \includegraphics[width=8.6cm]{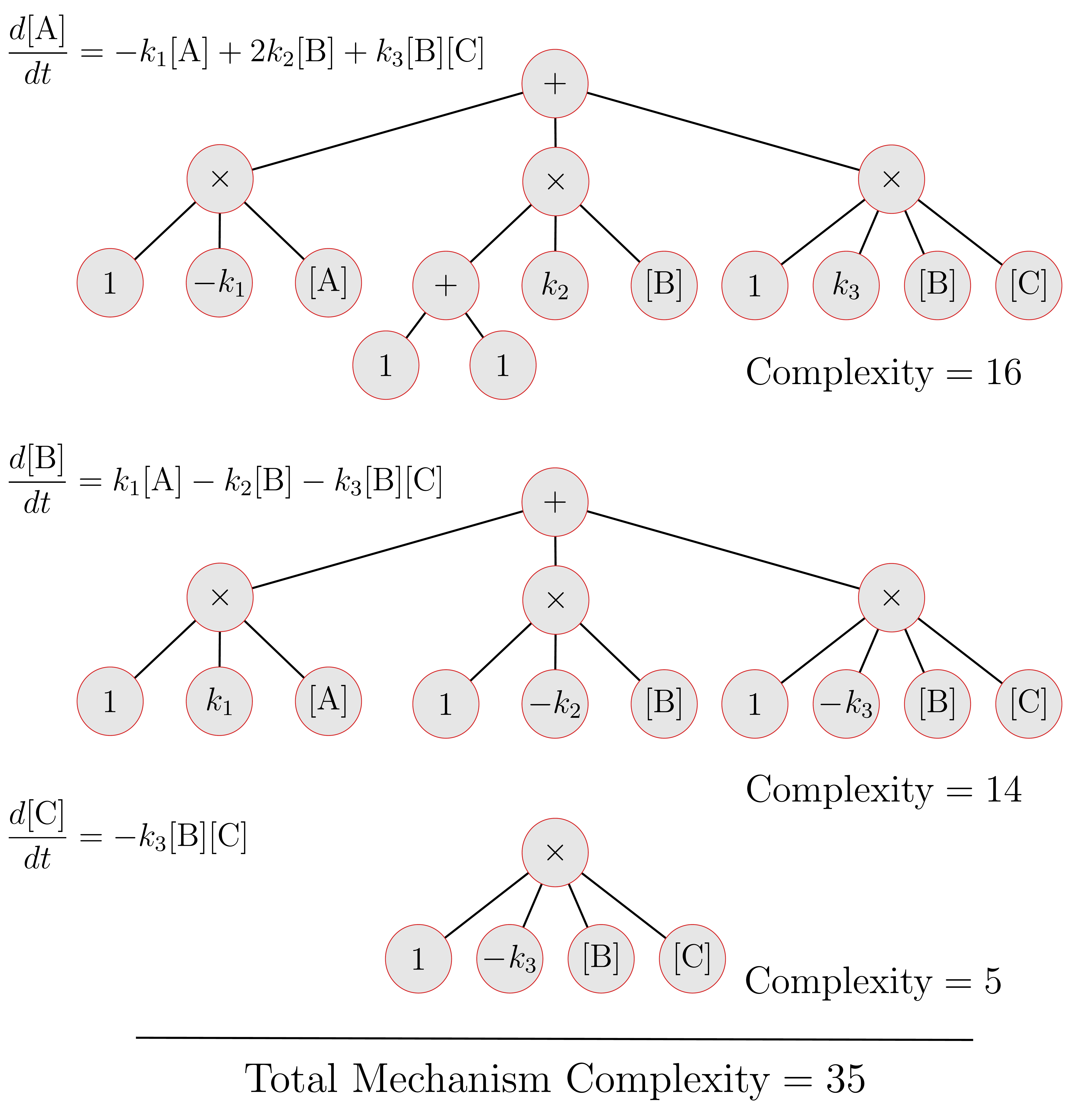}
 \caption{Expression tree complexity analysis for the example mechanism shown in Eq.~(\ref{eq:example2})}.
 \label{fig:tree}
\end{figure}

The complexity metric applied in this work arises from counting the nodes in an expression tree that represents
the kinetic equations for a mechanism (see Eq.~(\ref{eq:kinetic})).
Symbolic expression trees represent analytical functions in a hierarchical structure, where each node represents an operation or operand.
The overall complexity metric is a sum over the number of nodes in the expression tree for each species in a mechanism:
\begin{equation}
\text{Complexity}(\textbf{M}) = \sum^N_{i=1}|T_i|,
\end{equation}
where $T_i$ is the expression tree (the graph) of the kinetic equation for the $i$-th species and $|T_i|$ is the number of nodes in that tree.
Example expression trees are shown in Fig.~\ref{fig:tree}.

Several modifications to typical expression tree construction are implemented:
\begin{enumerate}
\item We want larger stoichiometric coefficients to contribute more to the complexity in comparison to smaller coefficients. Therefore, all stoichiometric coefficients are written as a separate node or subtree. If the stoichiometric coefficient $s = 1$, 
it is written as a single node. 
If the stoichiometric coefficient $s>1$, it is expressed as a subtree with $s+1$ nodes where the additional node is due to the ``+'' operation.
\item The ``power'' operation is not used to write nonlinear terms. Instead we write nonlinear terms such as $[\text{A}]^2$ using the product operation. This is because terms like $[\text{A}]^2$ should contribute the same complexity as terms like $[\text{A}][\text{B}]$.
\end{enumerate}

In general, chemical reaction mechanisms have specific mathematical forms that contain polynomials but do not include other types of functions such as trigonometric or exponential functions. This simplifies the complexity calculation because we do not have to  decided how to weight these different functions in the complexity hierarchy. 

To illustrate how the complexity metric is implemented, consider the example mechanism
\begin{equation}
\label{eq:example2}
\begin{aligned}
\ce{A &->[$k_1$] B} \\[0ex]
\ce{B &->[$k_2$] A + A} \\[0ex]
\ce{B +C &->[$k_3$] A} \\ 
\end{aligned}
\end{equation}
that is described by the set of kinetic equations 
\begin{equation}
\begin{aligned}
\label{eq:examplemech}
\frac{d[\text{A}]}{dt} &= - k_1 [\text{A}] + 2 k_2 [\text{B}]  + k_3 [\text{B}] [\text{C}] \\[0ex]
\frac{d[\text{B}]}{dt} &=   k_1 [\text{A}] - k_2 [\text{B}]  - k_3 [\text{B}] [\text{C}] \\[0ex]
\frac{d[\text{C}]}{dt} &= - k_3 [\text{B}] [\text{C}]  \\[0ex]
\end{aligned}
\end{equation}
Figure~\ref{fig:tree} illustrates the expression trees for each of the three equations in this system.
Three key points are:
\begin{enumerate}
\item In the subtree for the term $2 k_2 [\text{B}]$ in the $\frac{d[\text{A}]}{dt}$ equation, note that the stoichiometric coefficient is broken into
a sum. This adds a penalty on the stoichiometric values in the expression complexity.
\item Nonlinear terms such as $k_1 [\text{A}]^2$ contribute more complexity than linear terms such as $k_1 [\text{A}]$. This agrees with physical intuition that nonlinear expressions are more complex than linear expressions.
\item All rate constants contribute the same to the complexity. Meaning no penalty is placed on the values of the rate constants. This is to avoid problems with coefficient thresholding that can arise in other methods.
\end{enumerate}

The overall goal of the SISR procedure is to solve the multiobjective optimization problem
\begin{equation}
\underset{\mathbf{M},\mathbf{k}}{\mathrm{min}} (\mathcal{L}_\text{c},\mathrm{Complexity}),
\end{equation}
meaning we want to find the mechanism and the corresponding rate constant values that minimize the concentration error with respect to the ground truth data and also minimizes the complexity.

\section{Results \label{sec:results}}

\subsection{Sequential Linear Mechanism}
The first mechanism we examine is paradigmatic sequential linear mechanism 
\begin{equation}
\begin{aligned}
\ce{A ->[$k_1$] B} \\[0ex]
\ce{B ->[$k_2$] C} \\[0ex]
\ce{C ->[$k_3$] D} \\ 
\end{aligned}
\end{equation}
that is described by the set of kinetic equations for the time evolution of the concentrations: 
\begin{equation}
\begin{aligned}
\label{eq:seqmech}
\frac{d[\text{A}]}{dt} &= - k_1 [\text{A}] \\[0ex]
\frac{d[\text{B}]}{dt} &=  k_1 [\text{A}] - k_2 [\text{B}]  \\[0ex]
\frac{d[\text{C}]}{dt} &= k_2 [\text{B}] - k_3 [\text{C}]  \\[0ex]
\frac{d[\text{D}]}{dt} &= k_3 [\text{C}]  
\end{aligned}
\end{equation}
with rate constants $k_1 = 6.312\times10^{-5}\, \text{s}^{-1}$, $k_2 = 1.262\times10^{-4}\, \text{s}^{-1}$, and $k_3 = 3.156\times10^{-5}\,\text{s}^{-1}$
and concentrations given in millimolar.
The search space involves 142 reactions that can be combined into $\sim1.1\times 10^{10}$ possible mechanisms considering 
mechanisms with $2-6$ reactions. So the search space is large.
The kinetic equations were integrated using the Explicit Runge-Kutta method. We used this integration method for all examples in this manuscript. 500 equally spaced data points over the time interval $[0,100000]$ in units of seconds was fed into the SISR algorithm. 
For this mechanism, the SISR method was evolved for 10 generations over 5 islands with number of reactions $|\bold{M}| = 2-6$. A population size of 2000 was used for the $|\bold{M}| = 3-6$ islands while 
a population size of 500 was used for $|\bold{M}| = 2$ island because of the smaller number of possible mechanisms on that island. The maximum reaction order was $O=2$. 
The mutation rate was $\mathcal{M} = 0.1$. The elitism was $\mathcal{E} = 0.1$ (meaning the top $10\%$ of the solutions were retained across generations) except for the $|\bold{M}| = 2$ island where the elitism was $\mathcal{E} = 0.4$ due to the smaller population size. 

\begin{figure}
 \centering
 \includegraphics[width=8.6cm]{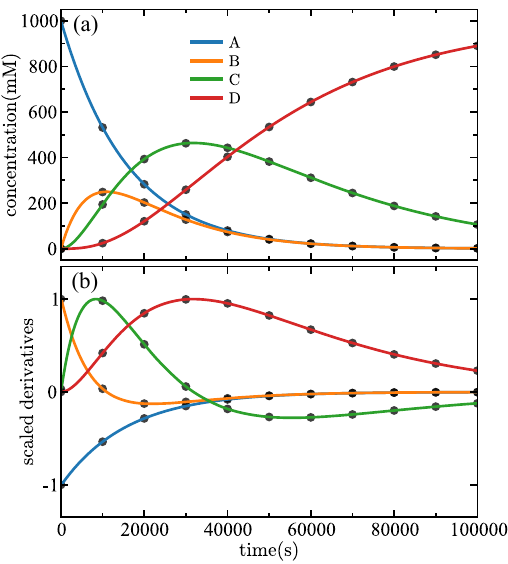}
 \caption{Time evolution of the (a) concentrations and (b) scaled derivatives---Eq.~(\ref{eq:ScaledDerivative})---%
 of each species in the sequential linear mechanism given in Eq.~(\ref{eq:seqmech}). The solid lines are the results of the SISR method and the corresponding black markers are a subset of the data used by SISR to extract the reaction mechanism and fit the rate constants. The concentrations are shown in units of millimolar and time is shown in units of seconds.}
 \label{fig:sequentialfit}
\end{figure}

The symbolic reaction extracted by SISR is:
\begin{equation}
\begin{aligned}
\frac{d\hat{[\text{A}]}}{dt} &= - \hat{k}_1 \hat{[\text{A}]} \\[0ex]
\frac{d\hat{[\text{B}]}}{dt} &=  \hat{k}_1 \hat{[\text{A}]}  - \hat{k}_2 \hat{[\text{B}]} \\[0ex]
\frac{d\hat{[\text{C}]}}{dt} &= \hat{k}_2 \hat{[\text{B}]} - \hat{k}_3 \hat{[\text{C}]}  \\[0ex]
\frac{d\hat{[\text{D}]}}{dt} &= \hat{k}_3 \hat{[\text{C}]}  
\end{aligned}
\end{equation}
which is exactly the true mechanism. 
The hat notation signifies that the rate constants and concentrations arise from the fits to the data.
The rate constants in the extracted reaction mechanism are $\hat{k}_1 = 6.310\times10^{-5}\, \text{s}^{-1}$, $\hat{k}_2 = 1.262\times10^{-4}\, \text{s}^{-1}$, and $\hat{k}_3 = 3.156\times10^{-5}\,\text{s}^{-1}$ in excellent agreement
(0.0317\% error or better) with the true rate constants. 
The results of the fitting are shown in Fig.~\ref{fig:sequentialfit}. The concentrations predicted by the SISR method 
closely match the time evolution of the true mechanism, 
as illustrated in Fig.~\ref{fig:sequentialfit}(a). In fact, at the presented level of visual fidelity, the ground truth data and the SISR result are indistinguishable. An important observation is that even though SISR searches over a symbolic space of nonlinear functions, the true linear mechanism is selected as the best model. Figure~\ref{fig:sequentialfit}(b) illustrates a comparison between ground truth and SISR results for the scaled numerical derivatives,
\begin{equation}
   \dfrac{\frac{d[\text{S}_i]} {dt}}
{\text{max}\left(\Big|\frac{d[\text{S}_i]}{dt}\Big|\right)}\;,  
   \label{eq:ScaledDerivative}
\end{equation}
demonstrating remarkable quantitative alignment. Note that the rate constants for each total reaction mechanism and the MSE used to sort mechanisms on each island are calculated in the derivative space. So, while the overall goal is to develop a mechanism that results in time series concentration profiles, analyzing the SISR results in the derivative space is illustrative of results for the fitting and sorting procedure.

We have found that for the linear sequential mechanism, the SISR search converges to the true symbolic mechanism using as few as 20 data points in the fitting procedure. 
This supports the possibility that SISR will perform well on sparse kinetic data, for example, on the types of data that could be generated in some experimental setups. 
As expected, the more data points that are used in the fitting the closer the extracted rate constants become to the true rate constants. It is also interesting to note that the fitting method gives good agreement even when a relatively small number of data points are used.

\begin{figure}
 \centering
 \includegraphics[width=8.6cm]{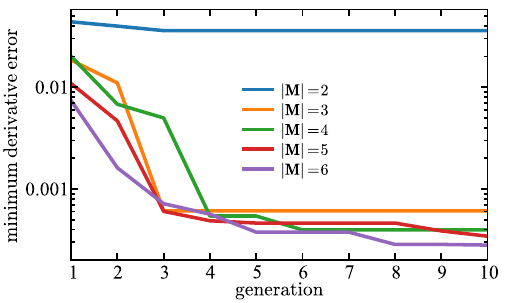}
 \caption{Minimum derivative error, \text{min}($\mathcal{L}_i$), as a function of generation using SISR to extract mechanisms from the ground truth data generated using the mechanism given in Eq.~(\ref{eq:seqmech}).
 The $y$-axis is shown on a log scale.
 Each curve is the result of a different island in the SISR method, where each island contains mechanisms with the number of reactions $|\bold{M}|$ shown in the legend.}
 \label{fig:minerror}
\end{figure}

Figure \ref{fig:minerror} illustrates how the minimum derivative error for each generation $i$, i.e.,  $\text{min}(\mathcal{L}_i)$,
changes over 10 generations of the SISR genetic algorithm. 
In this case, the data that was fed into the SISR method was concentration data from 50 equally spaced time points over the time interval $[0,100000]$. 
Each color curve in Fig.~\ref{fig:minerror}  represents the results for a different island size in the SISR algorithm.
Figure \ref{fig:minerror} demonstrates that the genetic evolution of mechanisms results in monotonically decreasing error over each generation.
This shows that the SISR algorithm is optimizing the symbolic mechanism and the genetic search tends toward an optimal mechanism on each island.
For $|\bold{M}| = 2$, only two reactions are included in each mechanism, and so the search space of possible mechanisms is small---approximately $10^4$  mechanisms.
Therefore, the optimal solution for that island is found after only a couple of generations because each generation includes a large portion of the total search space of possible mechanisms.
However, the error generated on the $|\bold{M}| = 2$ island is large compared to the other islands.
The ground truth mechanism has $|\bold{M}| = 3$ reactions, and there is a dramatic (approximately two orders of magnitude) drop in error when going from the 
$|\bold{M}| = 2$ island to the $|\bold{M}| = 3$ island. 
This illustrates an important result because to identify the correct mechanism a heuristic argument is to look
for the optimal mechanism on the island that occurs immediately after a steep drop in error with respect to variation in mechanism size.
Later we will give a more rigorous definition for mechanism selection, although we have found this heuristic observation to be an accurate identifier over the 
systems examined in this work.

In the derivative space, we see an interesting result which is that while the true mechanism has  $|\bold{M}| = 3$ reactions, the 
islands with $|\bold{M}| = 4$, $|\bold{M}| = 5$, and   $|\bold{M}| = 6$ reactions actually generate lower error than the true mechanism, 
although this is not the case in the concentration space as we will show next.
The cause of this reduction in error when adding spurious and erroneous reactions to the true mechanism size is that as more reactions are added, the algorithm has more coefficients to adjust in the fitting procedure, thereby fitting the data more closely. This means that the mechanisms on those islands are overfitting the data and doing so on symbolic functions that do not best fit the data.
This also illustrates the need for a complexity measure when making the final mechanism selection, because if only minimum error was used then the wrong mechanism would be chosen in this case.

\begin{figure}
 \centering
 \includegraphics[width=8.6cm]{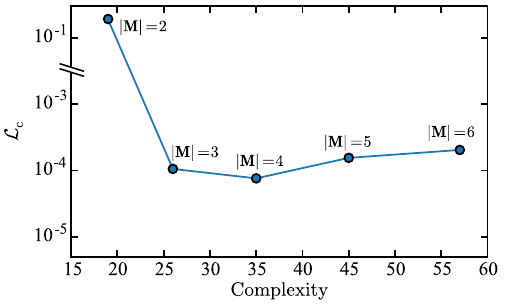}
 \caption{Complexity vs. concentration error $\mathcal{L}_\text{c}$ for the sequential linear mechanism. Each marker corresponds to the best mechanism as calculated using the derivative error $\mathcal{L}_\text{der}$ from the labeled island. The $y$-axis is shown on a log scale.}
 \label{fig:seqcomplex}
\end{figure}

Figure \ref{fig:seqcomplex} is a plot of the value for the loss function in the concentration space $\mathcal{L}_\text{c}$ 
as a function  of the mechanism complexity.
Each marker in the plot represents the mechanism with the lowest derivative error on each island. 
This Pareto front plot illustrates that the 
$|\bold{M}| = 3$ mechanism is the point on the Pareto front where increasing complexity results in limited improvement in the accuracy of the 
mechanism. Therefore the $|\bold{M}| = 3$ mechanism is the optimal solution.
This can be observed because there is a a steep drop in error when going from the $|\bold{M}| = 2$ point to the $|\bold{M}| = 3$ point,
and while the complexity for $|\bold{M}| = 2$ is smaller, the tradeoff in concentration error is too dramatic to make it the optimal solution.
When comparing the $|\bold{M}| = 3$ to $|\bold{M}| = 4$ mechanisms, there is limited improvement in concentration error when adding a new reaction, however
there is a large increase in complexity.
Interestingly, the $|\bold{M}| = 5$ and $|\bold{M}| = 6$ mechanisms result in an increase in error with respect to the $|\bold{M}| = 3$ mechanism. This illustrates that
including more terms in the symbolic mechanisms, i.e., including more reactions, does not necessarily result in improved accuracy but will generally result in 
a higher complexity.
The error metrics used to construct the Pareto front in Fig.~\ref{fig:seqcomplex} are shown in Table~\ref{tbl:seq} 
along with the corresponding derivative error 
for each mechanism.

\begin{table}[h]
\small
  \caption{\ SISR error metrics for the sequential linear mechanism on each island.}
  \label{tbl:seq}
  \begin{tabular*}{0.48\textwidth}{@{\extracolsep{\fill}}cccc}
    \hline
    $|\bold{M}|$ & $\mathcal{L}_\text{der}$ & $\mathcal{L}_\text{c}$ & \text{Complexity} \\
    \hline
    2 &$3.59\times10^{-2}$ & $1.88\times10^{-1}$  &19 \\
    3 & $6.09\times10^{-4}$& $1.05\times10^{-4}$  & 26\\
    4 & $3.97\times10^{-4}$& $7.58\times10^{-5}$ & 35 \\
    5 & $3.44\times10^{-4}$& $1.54\times10^{-4}$ & 45 \\
    6 & $2.81\times10^{-4}$& $2.03\times10^{-4}$ & 57 \\
    \hline
  \end{tabular*}
\end{table}

\subsubsection{Performance on Noisy Data}

\begin{figure}
 \centering
 \includegraphics[width=8.6cm]{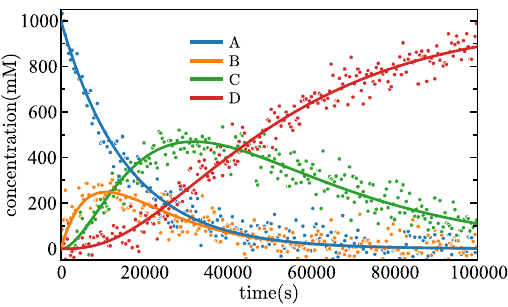}
 \caption{Time evolution of the concentrations  of each species in the sequential linear mechanism given in Eq.~(\ref{eq:seqmech}). The SISR fit for each species extracted from noisy data is shown by a solid curve.
 The markers are a subset of the noisy data used by SISR.}
 \label{fig:sequentialfitnoise}
\end{figure}

Noise can be a prominent feature in experiential and simulated chemical kinetics data, for example, in data obtained from molecular dynamics simulations using reactive force fields. Therefore, examining the robustness of the SISR approach in the presence of noise is an important metric. 
To this end, we examined the performance of SISR on noisy data by adding Gaussian noise sampled from the normal distribution $\mathcal{N} (0,50)$ to the same dataset used in the deterministic solutions of the sequential linear mechanism given in Eq.~(\ref{eq:seqmech}). A Savitzky-Golay (SG) filter was applied to the concentration data, and that filtered data was used as the SISR input. This is a general approach that has been that has worked previously due to known problems with numerical derivative calculations in the presence of noise \cite{de2020pysindy}. The result of the SISR on the noisy data is shown in Fig.~\ref{fig:sequentialfitnoise}, with excellent agreement observed between the noisy data and the mechanism predicted by SISR. 
The SISR method is able to find correct mechanism (the same as in Eq.\ref{eq:seqmech}). The rate constants extracted from the data were $\hat{k}_1 = 6.346\times10^{-5}\, \text{s}^{-1}$, $\hat{k}_2 = 1.279\times10^{-4}\, \text{s}^{-1}$, and $\hat{k}_3 = 3.091\times10^{-5}\,\text{s}^{-1}$ in strong agreement
with the true rate constants.  Note the specific rate constant values extracted from noisy data will depend on the specific realization of the noise.
An important point is that the SISR method is stochastic and not guaranteed to find the optimal solution (in this case meaning the correct mechanism) for each realization of the noise. Overall, this result illustrates the ability of SISR to extract the true reaction mechanism on data with high levels of noise.

\subsubsection{Hidden Variables and Intermediates}

\begin{figure}
 \centering
 \includegraphics[width=8.6cm]{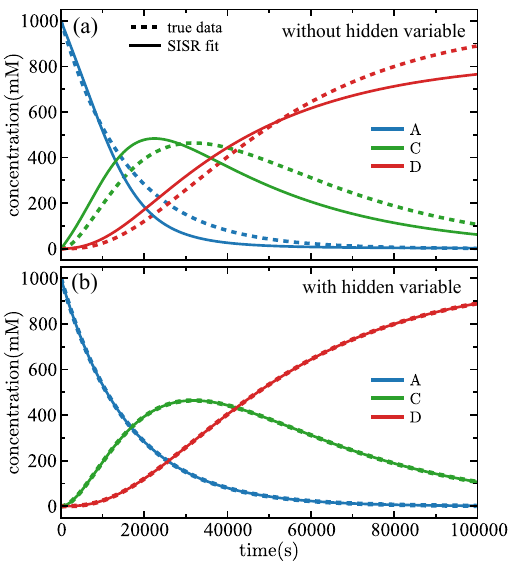}
 \caption{Time evolution of the concentrations of the species in the sequential linear mechanism given in Eq.~(\ref{eq:seqmech}). Panel (a) shows the case in which data for species A, C, and D are fed into SISR but no hidden variable (a new chemical species) is allowed in the mechanism and panel (b) shows the same case but with a hidden variable being allowed. In both cases, the error function only includes A, C, and D.
 The solid lines are the SISR fit and the dashed lines are the true data.}
 \label{fig:hidden}
\end{figure}

The SISR workflow can also be used to detect the presence of hidden variables (such as unknown chemical intermediates) in a set of chemical concentration data. 
The specific question we want to address is: Given a set of concentration data, does including more chemical species (like intermediates) in the symbolic regression part of the SISR algorithm beyond what is present in the data result in a more accurate reaction mechanism? 
To address this question, we consider the case of the sequential linear mechanism but where only a subset of the data is fed into SISR, for example concentration data for species A, B, and D or species A, C, and D. 
The SISR method is then used to detect the presence of the missing species (the hidden intermediate species).
In all cases considered in this section, the loss function is computed by only using species with available concentration data---the hidden intermediate is excluded from the error calculation.

Figure~\ref{fig:hidden}(a) is the SISR result for the case in which data from species A, C, and D are used (the unknown intermediate in this case is species B), and SISR is applied to find a mechanism containing only those species.
In this case, a poor fit and mechanism is obtained for the available data. The reason for this is that, without including the hidden intermediate, the model cannot reproduce the correct time evolution of the concentrations or reaction pathways that connect the observed species.
Now, compare those results with the results in Figure~\ref{fig:hidden}(b) where SISR is allowed to search for mechanisms that include one additional (previously hidden) species. In this case, the recovered mechanism correctly identifies the presence of the missing intermediate and yields a significantly improved fit to the data. Specifically, 
the error computed using Eq.~\ref{eq:MSE} drops by a factor of approx $10^5$ compared to the case shown in Figure~\ref{fig:hidden}(a) that does not include an intermediate in the reaction mechanism.
It is interesting to note that even though only data from three species is used, the SISR approach converges to the exact correct underlying mechanism. 
Therefore, including the intermediate species in the reaction mechanism, despite the absence of direct data for that species, yields a significantly improved mechanism. This demonstrates that the dataset implicitly contains evidence of a hidden intermediate, which is consistent with the known (see Eq.~(\ref{eq:seqmech})) underlying reaction mechanism.

We have also confirmed that a similar level of improvement is observed using hidden variables in the reaction mechanism when other species are removed from the data set. So, for example, when data for species A, B, and D are used with C being excluded and when data from species A, B, and C are used with D being excluded. In all cases we have studied for this mechanism, the application of SISR yields the exact true mechanism and also detects the presence of a hidden variable.
This proof-of-concept illustrates how SISR can be used to detect and then fit hidden chemical intermediates. Further work in this area will focus on constructing a multidimensional Pareto front that also includes the number of species involved in the process as an optimization dimension.

\subsection{Lotka-Volterra with Social Friction}

Next, we applied SISR to the Lotka-Volterra with Social Friction mechanism examined in Ref.~\citenum{ReactiveSINDY} using the Reactive SINDy method.
This mechanism exhibits the types of oscillatory behaviors seen in some biochemical systems \cite{bechtel2011complex}, for example in some viruses and in susceptive cells.
The specific mechanism is
\begin{equation}
\begin{aligned}
\label{eq:LVmech}
\ce{A + A &->[$k_1$] $\emptyset$} \\[0ex]
\ce{B + B &->[$k_2$] $\emptyset$} \\[0ex]
\ce{A  &->[$k_3$] A + A} \\[0ex]
\ce{A + B  &->[$k_4$] B + B} \\[0ex]
\ce{B &->[$k_5$] $\emptyset$} \\
\end{aligned}
\end{equation}
and involves two species $\text{A}$ and $\text{B}$. 
Where $\emptyset$ denotes the annihilation or irreversible removal of reactants.
The corresponding kinetic equations for the system are
\begin{equation}
\begin{aligned}
\label{eq:LVder}
\frac{d[\text{A}]}{dt} &= -2 k_1 [\text{A}]^2 + k_3 [\text{A}] - k_4[\text{A}][\text{B}]\\[0ex]
\frac{d[\text{B}]}{dt} &= -2 k_2 [\text{B}]^2 + k_4[\text{A}][\text{B}]- k_5 [\text{B}]\\
\end{aligned}
\end{equation}
with rate constant values given by $k_1 = 0.1 $, $k_2 = 0.1 $, $k_3 = 1$, $k_4 = 1$, $k_5 = 1$.

\begin{figure}
 \centering
 \includegraphics[width=8.6cm]{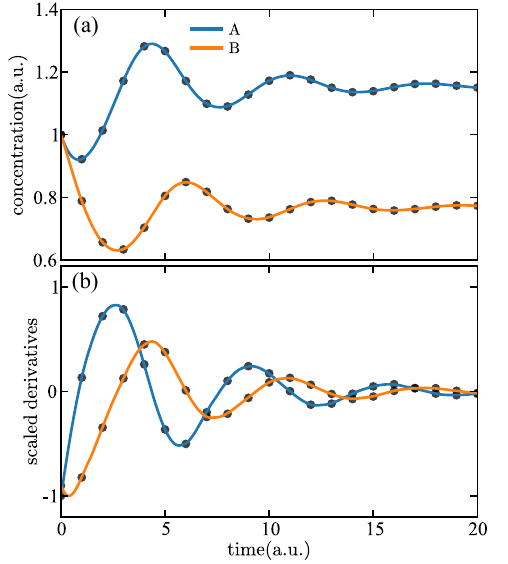}
 \caption{Time evolution of the (a) concentrations and (b) scaled derivatives---Eq.~(\ref{eq:LVmech})---of each species in the Lotka-Volterra mechanisms with social friction given in Eq.~(\ref{eq:LVder}). The solid lines are the results of the SISR method and the corresponding black markers are a subset of the data used by SISR to extract the reaction mechanism and fit the rate constants. The concentrations and time are shown in arbitrary units (a.u.).  }
 \label{fig:LVfit}
\end{figure}

To generate this mechanism, a population of mechanisms  was evolved for 20 generations over 6 islands with $|\bold{M}| = 2-7$ reactions. The population size was 2000 for 
the islands with $|\bold{M}| = 3-7$ reactions
and 500 for the $|\bold{M}| = 2$ island. The mutation rate was 0.1 and the elitism was 0.1 for all islands except the $|\bold{M}| = 2$ island where the elitism was 0.4 due to the smaller population size.  The maximum reaction order was $O=2$. The ground truth data was generated using 1000 equally-space points over the time interval $[0,20]$.
The search space involves 57 reactions that can be combined into $\sim3.1\times 10^8$ possible mechanisms considering 
mechanisms with  $2-7$ reactions. 

The model extracted from the data by the SISR method is
\begin{equation}
\begin{aligned}
\frac{d\hat{[\text{A}]}}{dt} &= -2 \hat{k}_1 \hat{[\text{A}]}^2 + \hat{k}_3 \hat{[\text{A}]} - \hat{k}_4\hat{[\text{A}]}\hat{[\text{B}]}\\[0ex]
\frac{d\hat{[\text{B}]}}{dt} &= -2 \hat{k}_2 \hat{[\text{B}]}^2 + \hat{k}_4\hat{[\text{A}]}\hat{[\text{B}]}- \hat{k}_5 \hat{[\text{B}]}\\
\end{aligned}
\end{equation}
which is, again, in exact agreement with the symbolic form for the ground truth reaction mechanism.
The values of the rate constants of this total mechanism were $\hat{k}_1 =  0.1003$, $\hat{k}_2 = 0.1029$, $\hat{k}_3 = 1.001$, $\hat{k}_4 = 1.001$, $\hat{k}_5 = 0.996$,
in strong agreement (2.81\% error or better) 
with the true rate constants.

The SISR results are shown in Fig.\ref{fig:LVfit} with panel (a) containing a comparison between the true and predicted concentration data and panel (b) containing the same comparison for the scaled derivative data. Excellent agreement is observed between the SISR result and the ground-truth data in both cases. The symbolic reaction mechanism is fitted on the derivative data, and Fig.\ref{fig:LVfit} (b) illustrates a primary advantage in using SISR: due to the inclusion of stoichiometric constraints, there is a distinct lack of overfitting the discovered mechanism, i.e., in the dynamical system that is extracted from the data.

\begin{figure}
 \centering
 \includegraphics[width=8.6cm]{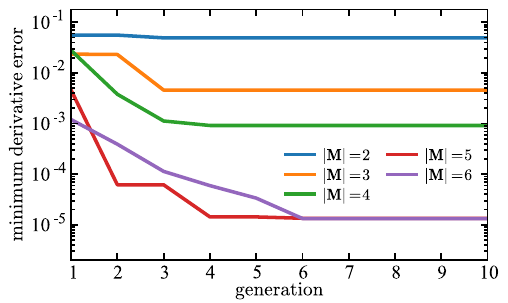}
 \caption{Minimum derivative error, \text{min}($\mathcal{L}_i$), as a function of generation using SISR to extract mechanisms from the ground truth data generated using the Lotka-Volterra mechanism given in Eq.~(\ref{eq:LVmech}).
 Each curve is the result of a different island in the SISR method, where each island contains mechanisms with the number of reactions $|\bold{M}|$ shown in the legend.}
 \label{fig:LVminerror}
\end{figure}

Figure \ref{fig:LVminerror} illustrates how the minimum derivative error changes over 10 generations when applying SISR to the Lotka-Volterra mechanism. 
Again, as in the previous mechanism, the genetic evolution results in monotonically decreasing error over each generation for every island size.
The error generated on the $|\bold{M}| = 2$ island is the largest.
As more reactions are added to the mechanism, the error decreases, which can be observed by comparing the results for each island size.
The ground truth mechanism has $|\bold{M}| = 5$ reactions, and the most dramatic (approximately two orders of magnitude) drop in error is observed when going from the $|\bold{M}| = 4$ island to the $|\bold{M}| = 5$ island. 
So for this reaction mechanism, the heuristic argument that the optimal mechanism occurs on the island that follows immediately after a steep drop in error with respect to variation in mechanism size would yield the selection of the true mechanism.
While the true mechanism has $|\bold{M}| = 5$ reactions, the 
island with $|\bold{M}| = 6$  reactions generates a lower error than $|\bold{M}| = 5$ island, although this difference is small ($1.35\times10^{-5}$ for $|\bold{M}| = 5$ compared to $1.33\times10^{-5}$ for $|\bold{M}| = 6$).
The reduction in error when adding spurious reactions to the mechanism is due to the algorithm having more coefficients to fit in the symbolic model.

\begin{figure}
 \centering
 \includegraphics[width=8.6cm]{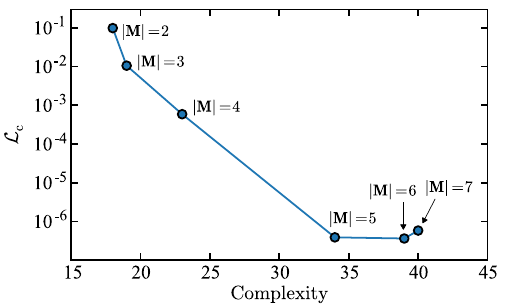}
 \caption{Complexity vs. concentration error $\mathcal{L}_\text{c}$ for the Lotka-Volterra mechanism. Each marker corresponds to the best mechanism as calculated using the derivative error $\mathcal{L}_\text{der}$ from the labeled island. The $y$-axis is shown on a log scale.}
 \label{fig:LVcomplex}
\end{figure}

Figure \ref{fig:LVcomplex} is a plot of the value for the loss function in the concentration space $\mathcal{L}_\text{c}$ 
as a function  of the mechanism complexity for the Lotka-Volterra mechanism.
The Pareto front shown in the plot illustrates that the
$|\bold{M}| = 5$ mechanism is the point where increasing complexity by adding more reactions results in limited improvement in the accuracy of the 
mechanism. Therefore the $|\bold{M}| = 5$ mechanism was chosen as the optimal solution.
There is a steep drop in error when going from the $|\bold{M}| = 2$ point to the $|\bold{M}| = 3$ point and from the $|\bold{M}| = 3$ point to the $|\bold{M}| = 4$. 
Comparing the $|\bold{M}| = 4$ point and the $|\bold{M}| = 5$ point (the true mechanism size) we see approximately a three orders of magnitude decrease in error. Therefore, despite the smaller mechanisms having less complexity, the tradeoff in increased error is too dramatic to make the smaller mechanisms the optimal solution. 
Compare this with the results for the $|\bold{M}| = 5$ and $|\bold{M}| = 6$ mechanisms where there is limited improvement in concentration error when adding a new reaction, but there is a significant increase in complexity.
The  $|\bold{M}| = 7$ mechanism shows an increase in error with respect to the $|\bold{M}| = 5$ and $|\bold{M}| = 6$ mechanisms.
The error metrics used to construct the Pareto front in Fig.~\ref{fig:LVcomplex} are shown in Table~\ref{tbl:LV} along with the 
corresponding derivative error for each mechanism.

\begin{table}[h]
\small
  \caption{\ SISR error metrics for the Lotka-Volterra mechanism on each island.}
  \label{tbl:LV}
  \begin{tabular*}{0.48\textwidth}{@{\extracolsep{\fill}}cccc}
    \hline
    $|\bold{M}|$ & $\mathcal{L}_\text{der}$ &  $\mathcal{L}_\text{c}$ & \text{Complexity} \\
    \hline
    2 & $4.93\times10^{-2}$ & $9.86\times10^{-2}$ &18\\
    3 & $4.58\times10^{-3}$& $1.06\times10^{-2}$ &19\\
    4 & $9.16\times10^{-4}$  & $5.89\times10^{-4}$ &23\\
    5 & $1.35\times10^{-5}$  & $3.88\times10^{-7}$ &34\\
    6 & $1.33\times10^{-5}$  & $3.62\times10^{-7}$ &39\\
    7 & $1.32\times10^{-5}$  & $5.82\times10^{-7}$ &40\\
    \hline
  \end{tabular*}
\end{table}

\subsubsection{Comparison to SINDy}
To illustrate how the SISR method compares to the well-used SINDy approach, 
we applied SINDy to the Lotka-Volterra mechanism.
Applying SINDy using the SR3 (sparse relaxed regularized regression)  method \cite{de2020pysindy, SR32019} with a threshold value of 0.01 resulted in 
\begin{equation}
\begin{aligned}
\frac{d\hat{[\text{A}]}}{dt} &= 
1.030 \hat{[\text{A}]} 
-  0.113 \hat{[\text{B}]}
-0.227 \hat{[\text{A}]}^2 \\[0ex]
& \quad 
+0.050 \hat{[\text{B}]}^2  -0.934\hat{[\text{A}]}\hat{[\text{B}]}\\[0ex]
\frac{d\hat{[\text{B}]}}{dt} &=
-0.224\hat{[\text{A}]} 
-  0.630 \hat{[\text{B}]}
+0.065 \hat{[\text{A}]}^2 \\[0ex]
& \quad -0.501 \hat{[\text{B}]}^2  +1.073\hat{[\text{A}]}\hat{[\text{B}]}\\
\end{aligned}
\end{equation}
which illustrates a principal problem that can arise when using the SINDy method without any physical constraints---overfitting in the derivative space.
Notice that every possible term in the symbolic dynamical system up to second-order has a non-zero coefficient and that, correspondingly, the discovered system is not in agreement with the true mechanism.
Increasing the SINDy threshold to a value of 0.25 in order to include fewer terms in the mechanism, i.e., to promote SINDy producing a sparser reaction mechanism, resulted in 
\begin{equation}
\begin{aligned}
\frac{d\hat{[\text{A}]}}{dt} &= 0.707 \hat{[\text{A}]} - 0.918\hat{[\text{A}]}\hat{[\text{B}]}\\[0ex]
\frac{d\hat{[\text{B}]}}{dt} &= - 0.951 \hat{[\text{B}]} -0.239 \hat{[\text{B}]}^2 + 0.983 \hat{[\text{A}]}\hat{[\text{B}]}\\
\end{aligned}
\end{equation}
which is again not in agreement with the true mechanism. 

Applying the Reactive SINDy method will result in similar problems to regular SINDy unless hyperparameters are tuned for the specific mechanism and a suitable basis set of candidate reactions are constructed.
Although it should be noted that after performing these tuning and construction tasks, Reactive SINDy generates an overall mechanism (reactions and rate constants) in excellent agreement with the true Lotka-Volterra mechanism \cite{ReactiveSINDY}.

\subsubsection{Performance on Noisy Data}

\begin{figure}
 \centering
 \includegraphics[width=8.6cm]{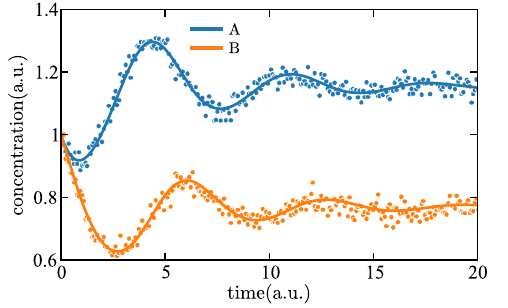}
 \caption{Time evolution of the concentrations of each species in the Lotka-Volterra mechanism given in Eq.~(\ref{eq:LVder}). 
 The SISR fit for each species extracted from noisy data is shown by a solid curve. The markers are a subset of the noisy data used in the SISR process.}
 \label{fig:LVnoise}
\end{figure}

We also examined the performance of SISR on the Lotka-Volterra mechanism with noisy concentration data. The results are shown in Fig.\ref{fig:LVnoise}. To generate the noisy data, Gaussian noise sampled from the normal distribution $\mathcal{N} (0,0.02)$ was added to the deterministic solutions of the system in Eq.~(\ref{eq:LVder}). We then applied a SG filter to the concentration data. The result of using this data as input to SISR is shown in Fig.~\ref{fig:LVnoise}, with excellent agreement observed between the noisy data and the mechanism predicted by SISR. This illustrates the ability of SISR to perform on concentration data with noise on a oscillatory reaction, a common situation in biological systems.

\subsection{Nonlinear Mechanism with Fast/Slow Dynamics}
\begin{figure}
 \centering
 \includegraphics[width=8.6cm]{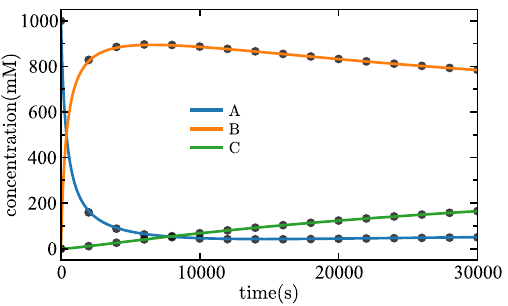}
 \caption{Time evolution of the concentrations of each species in the nonlinear mechanism given in Eq.~(\ref{eq:nonlinearmech}). The solid lines are the results of the SISR method and the corresponding black markers are a subset of the data used by SISR to extract the reaction mechanism and fit the rate constants. The concentrations are shown in units of millimolar and time is shown in units of seconds.}
 \label{fig:nonlinearfit}
\end{figure}

In order to test a nonlinear mechanism with fast/slow dynamics, i.e., a mechanism that has processes that evolve over disparate timescales, 
we examine the mechanism 
\begin{equation}
\begin{aligned}
\ce{A + A &->[$k_1$] B + B} \\[0ex]
\ce{B &->[$k_2$] C} \\[0ex]
\ce{B + C &->[$k_3$] A + A} \\ 
\end{aligned}
\end{equation}
The kinetic equations for this system are
\begin{equation}
\begin{aligned}
\label{eq:nonlinearmech}
\frac{d[\text{A}]}{dt} &= -2 k_1 [\text{A}]^2 + 2k_3 [\text{B}][\text{C}]\\[0ex]
\frac{d[\text{B}]}{dt} &= 2 k_1 [\text{A}]^2 - k_2 [\text{B}] - k_3 [\text{B}][\text{C}]\\[0ex]
\frac{d[\text{C}]}{dt} &=  k_2 [\text{B}] - k_3 [\text{B}][\text{C}]\\
\end{aligned}
\end{equation}
with rate constants $k_1 = 1.319\times10^{-6}\, \text{mM}^{-1} \text{s}^{-1}$, $k_2 = 9.125\times10^{-6}\, \text{s}^{-1}$, and $k_3 = 2.756\times10^{-8}\,\text{mM}^{-1} \text{s}^{-1}$
and concentrations being given in millimolar.
The fast/slow dynamics are obtained because the numerical value of the reaction rate $k_3$ is orders of magnitude different than the other
rate constants. 
In this mechanism, there is a rapid decline of $[\text{A}]$ and the corresponding rapid increase in $[\text{B}]$ (the fast processes) followed by a slow increase in  $[\text{C}]$ (the slow process), as shown in Fig.~\ref{fig:nonlinearfit}.

 The data fed into SISR was 2000 equally spaced time points over the interval $[0,20000]$ in units of seconds. 
For this mechanism, the search space involves 54 reactions that can be combined into 
$\sim2.9\times 10^7$ 
mechanisms. Respective population sizes of 2000 and 500 were evolved for 20 generations using islands with $|\bold{M}| = 3-5$ and $|\bold{M}| = 2$ reactions. The mutation rate was $\mathcal{M} = 0.1$ and the elitism was $\mathcal{E} = 0.1$.
The maximum reaction order was $O=2$. 
The best fit model found by the SISR method is
\begin{equation}
\begin{aligned}
\frac{d\hat{[\text{A}]}}{dt} &= -2 \hat{k}_1 \hat{[\text{A}]}^2 + 2\hat{k}_3 \hat{[\text{B}]}\hat{[\text{C}]}\\[0ex]
\frac{d\hat{[\text{B}]}}{dt} &= 2 \hat{k}_1 \hat{[\text{A}]}^2 - \hat{k}_2 \hat{[\text{B}]} - \hat{k}_3 \hat{[\text{B}]}\hat{[\text{C}]}\\[0ex]
\frac{d\hat{[\text{C}]}}{dt} &=  \hat{k}_2 \hat{[\text{B}]} - \hat{k}_3 \hat{[\text{B}]}\hat{[\text{C}]}\\
\end{aligned}
\end{equation}
Again, in exact agreement with the true mechanism. 
This mechanism was selected from the Pareto front in the Complexity vs. $\mathcal{L}_\text{c}$ space using the same multiobjective procedure described previously.																																								  
The rate constants in the found reaction mechanism are $\hat{k}_1 = 1.261\times10^{-6}\, \text{mM}^{-1} \text{s}^{-1}$, $\hat{k}_2 = 9.121\times10^{-6}\, \text{s}^{-1}$, and $\hat{k}_3 = 2.750\times10^{-8}\,\text{mM}^{-1} \text{s}^{-1}$ which very closely match the true rate constants (4.39\% error or better). 
The results of the SISR model are shown in Fig.~\ref{fig:nonlinearfit} with excellent agreement observed between the concentrations predicted by the SISR method and the concentrations calculated using the true mechanism. For times above 20000, the SISR model is forecasting, meaning no training data was used from that time interval. Excellent agreement is observed in the forecasted region.

\subsubsection{Comparison to SINDy}
\begin{figure}
 \centering
 \includegraphics[width=8.6cm]{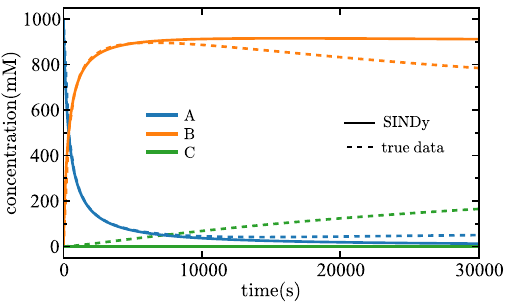}
 \caption{Time evolution of the concentrations of each species in the nonlinear mechanism given in Eq.~(\ref{eq:nonlinearmech}). The solid lines are the results of the SINDy method and the dashed lines are the true data. The concentrations are shown in units of millimolar and time is shown in units of seconds.}
 \label{fig:nonlinearfitSINDy}
\end{figure}

To illustrate how the SISR method compares to SINDy, 
we applied SINDy using SR3 optimization with a threshold of $10^{-8}$.
The data that was fed into SINDy was from 5000 equally spaced time points over the interval $[0,50000]$.
This resulted in a set of kinetic equations:
\begin{equation}
\begin{aligned}
\label{eq:nonlinearmechSINDY}
\frac{d\hat{[\text{A}]}}{dt} &= -2.64 \times 10^{-6} \hat{[\text{A}]}^2 + 5.52 \times 10^{-8}  \hat{[\text{B}]}\hat{[\text{C}]}\\[0ex]
\frac{d\hat{[\text{B}]}}{dt} &= 2.66 \times 10^{-6}\hat{[\text{A}]}^2
- 6.66 \times 10^{-8} \hat{[\text{A}]}\hat{[\text{B}]} \\
& \quad 
+ 1.27 \times 10^{-6}\hat{[\text{A}]}\hat{[\text{C}]}
- 1.49 \times 10^{-7}\hat{[\text{B}]}\hat{[\text{C}]}\\[0ex]
\frac{d\hat{[\text{C}]}}{dt} &=   3.01 \times 10^{-8} \hat{[\text{B}]}\hat{[\text{C}]}\\
\end{aligned}
\end{equation}
which illustrates several problems: (a) overfitting in the derivative space, (b) a lack of stoichiometric information,  and (c) the misidentification of chemical processes. 
To illustrate problem (c), note that the true mechanism has a process in the $\frac{d[\text{B}]}{dt}$ and $\frac{d[\text{C}]}{dt}$ equations involving a first-order reaction in species B, but the SINDy result does not.
Figure~\ref{fig:nonlinearfitSINDy} illustrates a comparison between the true data and the dynamical system derived using SINDy.
Notice that the slow rise of species C is not captured in the SINDy model. 
Compare this with the SISR result in  Fig.~\ref{fig:nonlinearfit} where that slow process is well captured.
It should be noted that  SINDy was not developed for chemical reaction mechanisms and is overall agnostic to the specific physics of a system. 
Although this can be a strength in certain situations, it hinders the discovery of accurate chemical reaction mechanisms.  The SISR method, however, encodes
physical information about chemical reactions and stoichiometry and therefore performs better with respect to accuracy on the selected examples in this work.

\subsection{Michaelis-Menten Kinetics}

The final example we examine is the Michaelis-Menten (MM) kinetic model, an important mechanism in biochemistry \cite{johnson2011original, cornish2015one}.
The specific MM model we use is the traditional form
\begin{equation}
\begin{aligned}
\ce{E + S &->[$k_1$] ES} \\[0ex]
\ce{ES &->[$k_2$] E + S} \\[0ex]
\ce{ES &->[$k_3$] E + P} \\ 
\end{aligned}
\end{equation}
with the corresponding set of kinetic equations
\begin{equation}
\begin{aligned}
\label{eq:MMmech}
\frac{d[\text{E}]}{dt} &= - k_1 [\text{E}][\text{S}] + k_2[\text{ES}] + k_3[\text{ES}]\\[0ex]
\frac{d[\text{S}]}{dt} &= - k_1 [\text{E}][\text{S}] + k_2[\text{ES}] \\[0ex]
\frac{d[\text{ES}]}{dt} &= k_1 [\text{E}][\text{S}]  - k_2[\text{ES}] - k_3[\text{ES}]  \\[0ex]
\frac{d[\text{P}]}{dt} &=  k_3 [\text{ES}]
\end{aligned}
\end{equation}
where E is an enzyme, S is a substrate, ES is an enzyme-substrate complex, and P is a product.
The rate constants used are $k_1 = 1\, \text{mM}^{-1} \text{s}^{-1}$, $k_2 = 0.1\, \text{s}^{-1}$, and $k_3 = 1\, \text{s}^{-1}$
The concentrations are given in millimolar and time is given in seconds. 

The data fed into SISR was 400 equally spaced time points over the interval $[0,2]$ in units of seconds. 
For this mechanism, the search space involves 182 reactions. A population size of 4000 was evolved for 20 generations for  islands with $|\bold{M}| = 3-5$ reactions and a population size of 500 was used for the $|\bold{M}| = 2$ island. The mutation rate was $\mathcal{M} = 0.1$ and the
elitism was $\mathcal{E} = 0.1$.
The population size for the $|\bold{M}| = 3-5$ islands was doubled from previous examples due to the larger number of possible reactions.
The maximum reaction order was $O=2$. 
The SISR result is
\begin{equation}
\begin{aligned}
\label{eq:MMmechSISR}
\frac{d\hat{[\text{E}]}}{dt} &= - \hat{k}_1 \hat{[\text{E}]}\hat{[\text{S}]} + \hat{k}_2\hat{[\text{ES}]} + \hat{k}_3\hat{[\text{ES}]}\\[0ex]
\frac{d\hat{[\text{S}]}}{dt} &= - \hat{k}_1 \hat{[\text{E}]}\hat{[\text{S}]} + \hat{k}_2\hat{[\text{ES}]} \\[0ex]
\frac{d\hat{[\text{ES}]}}{dt} &= \hat{k}_1 \hat{[\text{E}]}\hat{[\text{S}]}  - \hat{k}_2\hat{[\text{ES}]} - \hat{k}_3\hat{[\text{ES}]}  \\[0ex]
\frac{d\hat{[\text{P}]}}{dt} &=  \hat{k}_3 \hat{[\text{ES}]}
\end{aligned}
\end{equation}
which agrees exactly with the true mechanism.
The SISR mechanism was selected using the previously described procedure by comparing the error and complexity of the best performing mechanism from each island
and making a determination based on the shape of the Pareto front, specifically looking for the steepest drop in error which occurred between the $|\bold{M}| = 2$ and $|\bold{M}| = 3$ islands. Note that the $|\bold{M}| = 4$ island produced the lowest concentration error, but, as in the previous examined mechanisms, comparing the complexity of the mechanisms and error in a multiobjective picture results in the $|\bold{M}| = 3$ mechanism being chosen as the best overall mechanism.																																																																																																																																																													   
The fit rate constants for the extracted mechanism are $\hat{k}_1 = 0.993\, \text{mM}^{-1} \text{s}^{-1}$, $\hat{k}_2 = 0.084\, \text{s}^{-1}$, and $\hat{k}_3 = 1.000\, \text{s}^{-1}$,
which all have less than 17.0\% error
compared to the true rate constants.

\begin{figure}
 \centering
 \includegraphics[width=8.6cm]{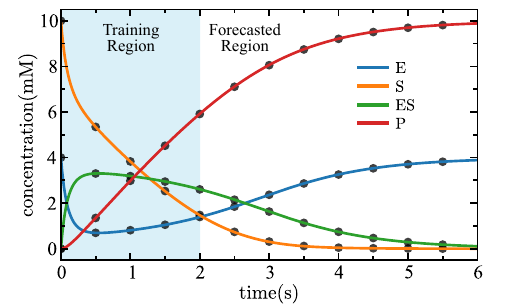}
 \caption{Time evolution of the concentrations of each species in the Michaelis-Menten   kinetic mechanism  given in Eq.~(\ref{eq:MMmech}). The solid lines are the results of the SISR method and the corresponding black markers are a subset of the data used by SISR to extract the reaction mechanism and fit the rate constants. Data from the light blue region is used to train SISR while data from the white region is forecasted.  The concentrations are shown in units of millimolar and time is shown in units of seconds. }
 \label{fig:MMfit}
\end{figure}

The results of SISR on the MM model is shown in Fig.~\ref{fig:MMfit}, with excellent agreement observed between the SISR-predicted concentrations and the ground truth data. 
The dynamics of the MM model are complex and involve competing processes that occur over different time scales. There are fast processes such as the fast decay of species E and the corresponding fast rise of species ES, and also slow processes such as the rise of species E and P to the steady-state values. SISR captures these processes well.

The region shown in blue  in Fig.~\ref{fig:MMfit} is the training region---data from this region was used to extract the mechanism and fit the rate constants. 
The white region in Fig.~\ref{fig:MMfit} shows the SISR result on data that was not used for mechanism discovery. Data in this region can be used to validate the SISR mechanism on unseen data and to assess the capability of SISR for time-series forecasting of chemical concentrations.  
Excellent agreement is observed between the ground truth data and the forecasted concentrations. This illustrates one of the principal advantages of using SISR for reaction discovery---because of the stoichiometrically-informed construction of the reaction mechanism, it is able to accurately forecast concentrations at future time points while avoiding the overfitting and extrapolation errors that are common in black-box machine learning. 

\subsection{Glucose Oxidation}

The final reaction mechanism we examine with SISR is glucose oxidation (GO), an important biochemical process.
The specific GO mechanism we consider consists of the conversion from $\alpha\text{-}\text{glucose}$ to $\beta\text{-}\text{glucose}$, the reverse reaction involving $\beta\text{-}\text{glucose}$ to $\alpha\text{-}\text{glucose}$ conversion, 
and the glucose oxidase (E) catalyzed reaction of $\beta\text{-}\text{glucose}$ with oxygen to produce another glucose molecule---$\delta\text{-}\text{glucose}$---and $\text{H}_2 \text{O}_2$.
The individual reactions for this model are 
\begin{equation}
\begin{aligned}
\label{eq:GOmech}
\ce{$\alpha$\text{-}glucose &->[$k_1$] $\beta$\text{-}glucose} \\[0ex]
\ce{$\beta$\text{-}glucose &->[$k_{-1}$] $\alpha$\text{-}glucose} \\[0ex]
\ce{$\beta$\text{-}glucose + O_2 + E &->[$k_2$] $\delta$\text{-}glucose + H_2 O_2 + E}  \\ 
\end{aligned}
\end{equation}
which give rise to the corresponding set of kinetic equations
\begin{equation}
\begin{aligned}
\label{eq:GOmechkinetic}
\frac{d[\alpha]}{dt} &= - k_1 [\alpha] + k_{-1}[\beta]\\[0ex]
\frac{d[\beta]}{dt} &= k_1 [\alpha] - k_{-1}[\beta] - k_2 [\beta][\text{O}_2]  \\[0ex]
\frac{d[\delta]}{dt} &= k_2 [\beta][\text{O}_2] [\text{E}]  \\[0ex]
\frac{d[\text{E}]}{dt} &= 0 \\[0ex]
\frac{d[\text{O}_2]}{dt} &= - k_2 [\beta][\text{O}_2] [\text{E}] \\[0ex]
\frac{d[\text{H}_2 \text{O}_2]}{dt} &=  k_2 [\beta][\text{O}_2] [\text{E}]
\end{aligned}
\end{equation}
We assume that the enzyme has a steady state concentration of $1\mu \text{M}$
and that we have \textit{a priori} knowledge of the enzyme interactions.
Two sets of rates constants are examined. The values in the first set are  $k_1 = 4.45\times10^{-4} \text{s}^{-1}$, $k_{-1} = 3.03\times10^{-4} \text{s}^{-1}$, and $k_2 = 2.78\times10^{-3}\,\mu\text{M}^{-2} \text{s}^{-1}$. In the second set,  the value of $k_2$ is increased by an order of magnitude to $k_2 = 2.78\times10^{-2}\,\mu\text{M}^{-2} \text{s}^{-1}$. These values are based on work in Ref.\citenum{tao2009kinetic}.

\begin{figure}
 \centering
 \includegraphics[width=8.6cm]{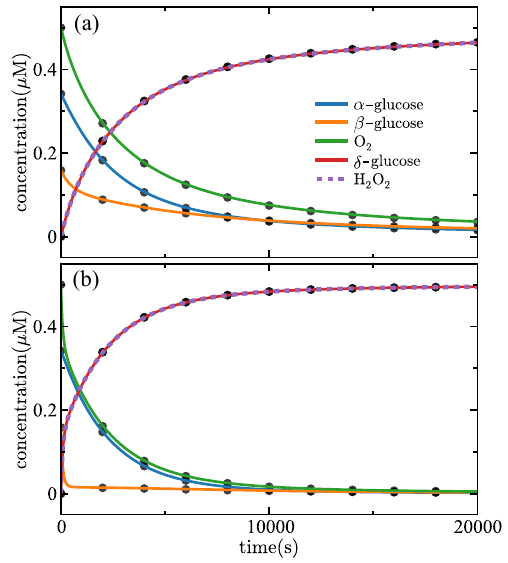}
 \caption{Time evolution of the concentrations of chemical species in the Glucose Oxidation reaction mechanism given in  Eq.~(\ref{eq:GOmech}). Panel (a) shows the case with $k_2 = 2.78\times10^{-3}\,\mu\text{M}^{-2} \text{s}^{-1}$  and panel (b) shows the case with $k_2 = 2.78\times10^{-4}\,\mu\text{M}^{-2} \text{s}^{-1}$. The solid lines are the results of the SISR method and the corresponding black markers are a subset of the data used by SISR to extract the reaction mechanism and fit the rate constants. The concentrations are shown in micromolar and time is shown in seconds}.
 \label{fig:GOfit}
\end{figure}

For the first set of rate constants, using SISR with a population size of 2000 and with 500 data points over the interval  $[0,10000]$,  SISR discovered the correct reactions and reaction mechanism:
\begin{equation}
\begin{aligned}
\label{eq:GOmech2}
\frac{d\hat{[\alpha]}}{dt} &= - \hat{k}_1 \hat{[\alpha]} + \hat{k}_{-1}\hat{[\beta]}\\[0ex]
\frac{d\hat{[\beta]}}{dt} &= \hat{k}_1 \hat{[\alpha]} - \hat{k}_{-1}\hat{[\beta]} - \hat{k}_2 \hat{[\beta]}\hat{[\text{O}_2]}  \\[0ex]
\frac{d\hat{[\delta]}}{dt} &= \hat{k}_2 \hat{[\beta]}\hat{[\text{O}_2]} \hat{[\text{E}]} \\[0ex]
\frac{d\hat{[\text{E}]}}{dt} &= 0 \\[0ex]
\frac{d\hat{[\text{O}_2]}}{dt} &= - \hat{k}_2 \hat{[\beta]}\hat{[\text{O}_2]} \hat{[\text{E}]}  \\[0ex]
\frac{d\hat{[\text{H}_2 \text{O}_2]}}{dt} &=  \hat{k}_2 \hat{[\beta]}\hat{[\text{O}_2]} \hat{[\text{E}]} 
\end{aligned}
\end{equation}
The extracted rate constants were  $\hat{k}_1 = 4.45\times10^{-4} \text{s}^{-1}$, $\hat{k}_{-1} = 3.23\times10^{-4} \text{s}^{-1}$, and $\hat{k}_2 = 2.73\times10^{-3}\,\mu\text{M}^{-2} \text{s}^{-1}$ in strong agreement with the true values.
A comparison between input data and the SISR fit is shown in Fig.~\ref{fig:GOfit}(a).

For the second set of rate constants, a population size of 4000 with 1000 data points over the time interval $[0,20000]$ in units of seconds was used.
Again, the exact reaction mechanism was found using SISR, illustrating the ability of SISR to account for fast/slow dynamics.
The extracted rate constants were  $\hat{k}_1 = 4.40\times10^{-4} \text{s}^{-1}$, $\hat{k}_{-1} = 2.56\times10^{-4} \text{s}^{-1}$, and $\hat{k}_2 = 2.63\times10^{-2}\,\mu\text{M}^{-2} \text{s}^{-1}$ in good agreement with the true values.
The result of the SISR fit is shown in Fig.~\ref{fig:GOfit}(b).

With respect to the performance of SISR on fast/slow systems, three points are of note:
(1) Compared to coefficient thresholding-based methods in the literature (such as SINDy-based approaches), SISR appears better able to extract fast/slow dynamics, (2) samples are taken in the fast region in the training data which is why the SISR method can recover the dynamics in that region, however inference in the absence of training data may capture these dynamics, and (3) further investigation on the use of SISR for fast/slow and stiff chemical kinetic systems is an important line of inquiry.

\section{Conclusions \label{sec:conc}}
A stoichiometrically-informed symbolic regression (SISR) method was introduced to automate the discovery of chemical reaction mechanisms from time series chemical concentration data. 
The SISR method generates symbolic forms for reaction mechanisms, and will be particularly useful when examining complex chemical systems where manual derivation of a mechanism is difficult or simply impractical.
By integrating a physics-informed mathematical framework that accounts for the intrinsic functional geometry of chemical reaction mechanisms and the stoichiometry of those mechanisms, the developed method successfully identified sparse and interpretable reaction mechanisms for multiple example chemical processes. 
Specifically, the SISR results demonstrated strong agreement between predicted and true mechanisms across a range of chemical reaction schemes.

Compared to existing approaches for symbolic dynamical system discovery such as SINDy and its variants \cite{Brunton2016SINDY,ReactiveSINDY, Bhatt2023SINDYCRN}, the SISR method alleviates key limitations by (a) incorporating stoichiometric constraints, (b) reducing the reliance on predefined reaction assumptions, and (c) improving robustness in handling fast-slow dynamics due to the reduction of hyperparameter tuning and thresholding procedures.
While SINDy-based approaches can offer significant computational advantages in comparison to SISR, for situations in which physical insight, interpretability, accuracy, and the ability to forecast future system states are important, SISR offers strong advantages.

 A key limitation of the current approach is that the rate constants extracted from the data are purely numerical values and not symbolic functional forms that can be applied to account for changes in environmental and/or thermodynamic conditions such as changes in temperature or pressure.  This can be a problem if the derived mechanism will be used to forecast the outcome of the chemical process at a thermodynamic state other than the state used to generate/measure the data. Additionally, due to this limitation, nonequilibrium cases in which the rate constants are varying in time as parameteric coefficients \cite{Li2019timevaryingSINDY, Rudy2019parametric, YangKress2023} or over thermodynamic conditions are currently outside the scope of the current approach. Future work will address nonequilibrium chemical reaction mechanisms \cite{Bazant2013, craven16a,hern21j}. Improving the computational efficiency of the method using different optimization strategies is an important next step. Additionally, validation across a broad set of experimental datasets would further establish the utility of SISR on noisy, incomplete, and complex data, and work in this direction is currently underway.



\section*{Acknowledgments}

We acknowledge support from the Los Alamos National Laboratory (LANL) Directed Research and Development funds (LDRD).
We also acknowledge support to RH and to MPB from the National Science Foundation grant No.~2102455.
The computing resources used to perform this
research were provided in part by the LANL Institutional Computing Program,
and by
the Advanced Research Computing at Hopkins
(ARCH) high-performance computing (HPC) facilities.




\bibliographystyle{apsrev}

\begin{thebibliography}{76}
\bibitem[{not()}]{note1}
\bibinfo{note}{The scaling factor is not applied in cases where the maximum concentration is zero.}
\end{thebibliography}

\end{document}